\documentclass[a4paper,11pt]{article}
\pdfoutput=1 
\usepackage{multirow}
\usepackage{soul}             
\usepackage{jheppub} 
\usepackage[T1]{fontenc} 

\newcommand\e[1]{10^{-#1}}

\newcommand\y[1]{\Upsilon(#1S)}

\newcommand\ppy{\y2\rightarrow\pi^+\pi^-\y1}

\newcommand\yee{\y1\rightarrow e^+e^-}
\newcommand\ymm{\y1\rightarrow\mu^+\mu^-}
\newcommand\ytt{\y1\rightarrow\tau^+\tau^-}

\newcommand\yll{\y1\rightarrow\ell^\pm\ell^{\prime\mp}}
\newcommand\Yll{\y1\rightarrow\ell^\pm\ell^\mp}
\newcommand\ygll{\y1\rightarrow\gamma\ell^\pm\ell^{\prime\mp}}
\newcommand\yem{\y1\rightarrow e^\pm\mu^\mp}
\newcommand\ymt{\y1\rightarrow \mu^\pm\tau^\mp}
\newcommand\yet{\y1\rightarrow e^\pm\tau^\mp}
\newcommand\ylt{\y1\rightarrow\ell^\pm\tau^\mp}
\newcommand\ygem{\y1\rightarrow \gamma e^\pm\mu^\mp}
\newcommand\ygmt{\y1\rightarrow \gamma \mu^\pm\tau^\mp}
\newcommand\yget{\y1\rightarrow \gamma e^\pm\tau^\mp}
\newcommand\yglt{\y1\rightarrow \gamma \ell^\pm\tau^\mp}

\newcommand\te{\tau^-\rightarrow e^-\bar{\nu}_e\nu_\tau}
\newcommand\tm{\tau^-\rightarrow \mu^-\bar{\nu}_\mu\nu_\tau}

\newcommand\tP{\tau^-\rightarrow\pi^-\pi^+\pi^-\nu_\tau}

\newcommand\mt[1]{\mathcal{#1}}

\newcommand\dm{\Delta M}
\newcommand\gev{{\rm GeV}/c^2}
\newcommand\mev{{\rm MeV}/c^2}
\newcommand\Gev{{\rm GeV}/c}
\newcommand\Mev{{\rm MeV}/c}
\newcommand\mr[1]{ M^{{\rm recoil}}_{\rm #1}}

\newcommand\fb{{\rm fb}^{-1}}
\newcommand\mrpp{\mr{\pi\pi}}
\newcommand\mrppm{\mr{\pi\pi\mu}}
\newcommand\mrppe{\mr{\pi\pi e}}
\newcommand\mrppl{\mr{\pi\pi\ell}}

\newcommand\mrppgl{\mr{\pi\pi\ell\gamma}}

\newcommand\pvt{p^{\tau}_{\rm vis}}
\newcommand\N[1]{N_{\rm #1}}
\newcommand\Nm{\N{\mu}}
\newcommand\Ne{\N{e}}

\newcommand\cospp{\cos \theta_{\pi\pi}}

\newcommand\m[1]{M_{\rm #1}}
\newcommand\E[1]{E_{\rm #1}}
\newcommand\p[1]{|{\overrightarrow p}_{#1}|}
\newcommand\CVL{C_{VL}^{q\ell\ell^\prime}}
\newcommand\CVR{C_{VR}^{q\ell\ell^\prime}}
\newcommand\CAL{C_{AL}^{q\ell\ell^\prime}}
\newcommand\CAR{C_{AR}^{q\ell\ell^\prime}}
\newcommand\CTR{C_{TR}^{q\ell\ell^\prime}}
\newcommand\CTL{C_{TL}^{q\ell\ell^\prime}}

\title{\boldmath Search for charged lepton flavor violating decays of $\y1$}

\newcounter{AffiliationCounter}
\stepcounter{AffiliationCounter}\edef\instBilbao{\protect\theAffiliationCounter}
\stepcounter{AffiliationCounter}\edef\instBonn{\protect\theAffiliationCounter}
\stepcounter{AffiliationCounter}\edef\instBNL{\protect\theAffiliationCounter}
\stepcounter{AffiliationCounter}\edef\instBINP{\protect\theAffiliationCounter}
\stepcounter{AffiliationCounter}\edef\instCharles{\protect\theAffiliationCounter}
\stepcounter{AffiliationCounter}\edef\instChonnam{\protect\theAffiliationCounter}
\stepcounter{AffiliationCounter}\edef\instCAU{\protect\theAffiliationCounter}
\stepcounter{AffiliationCounter}\edef\instCincinnati{\protect\theAffiliationCounter}
\stepcounter{AffiliationCounter}\edef\instDESY{\protect\theAffiliationCounter}
\stepcounter{AffiliationCounter}\edef\instDuke{\protect\theAffiliationCounter}
\stepcounter{AffiliationCounter}\edef\instDuyTan{\protect\theAffiliationCounter}
\stepcounter{AffiliationCounter}\edef\instFlorida{\protect\theAffiliationCounter}
\stepcounter{AffiliationCounter}\edef\instFuJen{\protect\theAffiliationCounter}
\stepcounter{AffiliationCounter}\edef\instFudan{\protect\theAffiliationCounter}
\stepcounter{AffiliationCounter}\edef\instGiessen{\protect\theAffiliationCounter}
\stepcounter{AffiliationCounter}\edef\instGifu{\protect\theAffiliationCounter}
\stepcounter{AffiliationCounter}\edef\instGoettingen{\protect\theAffiliationCounter}
\stepcounter{AffiliationCounter}\edef\instSokendai{\protect\theAffiliationCounter}
\stepcounter{AffiliationCounter}\edef\instGyeongsang{\protect\theAffiliationCounter}
\stepcounter{AffiliationCounter}\edef\instHanyang{\protect\theAffiliationCounter}
\stepcounter{AffiliationCounter}\edef\instHawaii{\protect\theAffiliationCounter}
\stepcounter{AffiliationCounter}\edef\instKEK{\protect\theAffiliationCounter}
\stepcounter{AffiliationCounter}\edef\instJPARC{\protect\theAffiliationCounter}
\stepcounter{AffiliationCounter}\edef\instHSE{\protect\theAffiliationCounter}
\stepcounter{AffiliationCounter}\edef\instJuelich{\protect\theAffiliationCounter}
\stepcounter{AffiliationCounter}\edef\instIKER{\protect\theAffiliationCounter}
\stepcounter{AffiliationCounter}\edef\instIISERM{\protect\theAffiliationCounter}
\stepcounter{AffiliationCounter}\edef\instIITB{\protect\theAffiliationCounter}
\stepcounter{AffiliationCounter}\edef\instIITG{\protect\theAffiliationCounter}
\stepcounter{AffiliationCounter}\edef\instIITH{\protect\theAffiliationCounter}
\stepcounter{AffiliationCounter}\edef\instIITM{\protect\theAffiliationCounter}
\stepcounter{AffiliationCounter}\edef\instIndiana{\protect\theAffiliationCounter}
\stepcounter{AffiliationCounter}\edef\instIHEP{\protect\theAffiliationCounter}
\stepcounter{AffiliationCounter}\edef\instProtvino{\protect\theAffiliationCounter}
\stepcounter{AffiliationCounter}\edef\instVienna{\protect\theAffiliationCounter}
\stepcounter{AffiliationCounter}\edef\instNapoli{\protect\theAffiliationCounter}
\stepcounter{AffiliationCounter}\edef\instRomaTre{\protect\theAffiliationCounter}
\stepcounter{AffiliationCounter}\edef\instTorino{\protect\theAffiliationCounter}
\stepcounter{AffiliationCounter}\edef\instISU{\protect\theAffiliationCounter}
\stepcounter{AffiliationCounter}\edef\instJAEA{\protect\theAffiliationCounter}
\stepcounter{AffiliationCounter}\edef\instJSI{\protect\theAffiliationCounter}
\stepcounter{AffiliationCounter}\edef\instKarlsruhe{\protect\theAffiliationCounter}
\stepcounter{AffiliationCounter}\edef\instIPMU{\protect\theAffiliationCounter}
\stepcounter{AffiliationCounter}\edef\instKAU{\protect\theAffiliationCounter}
\stepcounter{AffiliationCounter}\edef\instKitasato{\protect\theAffiliationCounter}
\stepcounter{AffiliationCounter}\edef\instKISTI{\protect\theAffiliationCounter}
\stepcounter{AffiliationCounter}\edef\instKorea{\protect\theAffiliationCounter}
\stepcounter{AffiliationCounter}\edef\instKyungpook{\protect\theAffiliationCounter}
\stepcounter{AffiliationCounter}\edef\instIJCLab{\protect\theAffiliationCounter}
\stepcounter{AffiliationCounter}\edef\instLebedev{\protect\theAffiliationCounter}
\stepcounter{AffiliationCounter}\edef\instLNNU{\protect\theAffiliationCounter}
\stepcounter{AffiliationCounter}\edef\instLjubljana{\protect\theAffiliationCounter}
\stepcounter{AffiliationCounter}\edef\instLMU{\protect\theAffiliationCounter}
\stepcounter{AffiliationCounter}\edef\instLuther{\protect\theAffiliationCounter}
\stepcounter{AffiliationCounter}\edef\instMNIT{\protect\theAffiliationCounter}
\stepcounter{AffiliationCounter}\edef\instMaribor{\protect\theAffiliationCounter}
\stepcounter{AffiliationCounter}\edef\instMPI{\protect\theAffiliationCounter}
\stepcounter{AffiliationCounter}\edef\instMelbourne{\protect\theAffiliationCounter}
\stepcounter{AffiliationCounter}\edef\instMississippi{\protect\theAffiliationCounter}
\stepcounter{AffiliationCounter}\edef\instMiyazaki{\protect\theAffiliationCounter}
\stepcounter{AffiliationCounter}\edef\instMEPhI{\protect\theAffiliationCounter}
\stepcounter{AffiliationCounter}\edef\instNagoya{\protect\theAffiliationCounter}
\stepcounter{AffiliationCounter}\edef\instNagoyaKMI{\protect\theAffiliationCounter}
\stepcounter{AffiliationCounter}\edef\instUNapoli{\protect\theAffiliationCounter}
\stepcounter{AffiliationCounter}\edef\instNara{\protect\theAffiliationCounter}
\stepcounter{AffiliationCounter}\edef\instNCU{\protect\theAffiliationCounter}
\stepcounter{AffiliationCounter}\edef\instNUU{\protect\theAffiliationCounter}
\stepcounter{AffiliationCounter}\edef\instTaiwan{\protect\theAffiliationCounter}
\stepcounter{AffiliationCounter}\edef\instKrakow{\protect\theAffiliationCounter}
\stepcounter{AffiliationCounter}\edef\instNihonDental{\protect\theAffiliationCounter}
\stepcounter{AffiliationCounter}\edef\instNiigata{\protect\theAffiliationCounter}
\stepcounter{AffiliationCounter}\edef\instNovosibirsk{\protect\theAffiliationCounter}
\stepcounter{AffiliationCounter}\edef\instOsakaCity{\protect\theAffiliationCounter}
\stepcounter{AffiliationCounter}\edef\instPNNL{\protect\theAffiliationCounter}
\stepcounter{AffiliationCounter}\edef\instPanjab{\protect\theAffiliationCounter}
\stepcounter{AffiliationCounter}\edef\instPittsburgh{\protect\theAffiliationCounter}
\stepcounter{AffiliationCounter}\edef\instPunjab{\protect\theAffiliationCounter}
\stepcounter{AffiliationCounter}\edef\instNPC{\protect\theAffiliationCounter}
\stepcounter{AffiliationCounter}\edef\instRIKENMSL{\protect\theAffiliationCounter}
\stepcounter{AffiliationCounter}\edef\instUSTC{\protect\theAffiliationCounter}
\stepcounter{AffiliationCounter}\edef\instShoyaku{\protect\theAffiliationCounter}
\stepcounter{AffiliationCounter}\edef\instSoochow{\protect\theAffiliationCounter}
\stepcounter{AffiliationCounter}\edef\instSoongsil{\protect\theAffiliationCounter}
\stepcounter{AffiliationCounter}\edef\instSungkyunkwan{\protect\theAffiliationCounter}
\stepcounter{AffiliationCounter}\edef\instSydney{\protect\theAffiliationCounter}
\stepcounter{AffiliationCounter}\edef\instTabuk{\protect\theAffiliationCounter}
\stepcounter{AffiliationCounter}\edef\instTata{\protect\theAffiliationCounter}
\stepcounter{AffiliationCounter}\edef\instTUM{\protect\theAffiliationCounter}
\stepcounter{AffiliationCounter}\edef\instTelAviv{\protect\theAffiliationCounter}
\stepcounter{AffiliationCounter}\edef\instToho{\protect\theAffiliationCounter}
\stepcounter{AffiliationCounter}\edef\instTohoku{\protect\theAffiliationCounter}
\stepcounter{AffiliationCounter}\edef\instERI{\protect\theAffiliationCounter}
\stepcounter{AffiliationCounter}\edef\instTokyo{\protect\theAffiliationCounter}
\stepcounter{AffiliationCounter}\edef\instTIT{\protect\theAffiliationCounter}
\stepcounter{AffiliationCounter}\edef\instTMU{\protect\theAffiliationCounter}
\stepcounter{AffiliationCounter}\edef\instVPI{\protect\theAffiliationCounter}
\stepcounter{AffiliationCounter}\edef\instWayneState{\protect\theAffiliationCounter}
\stepcounter{AffiliationCounter}\edef\instYamagata{\protect\theAffiliationCounter}
\stepcounter{AffiliationCounter}\edef\instYonsei{\protect\theAffiliationCounter}

\collaboration{The Belle Collaboration}
  \author[\instIISERM,\hbox{$\dagger$}]{S.~Patra, \note[$\dagger$]{Corresponding author.}} 
  \author[\instIISERM]{V.~Bhardwaj,} 
  \author[\instIJCLab]{K.~Trabelsi,} 
  \author[\instKEK,\instSokendai]{I.~Adachi,} 
  \author[\instTokyo]{H.~Aihara,} 
  \author[\instTabuk,\instKAU]{S.~Al~Said,} 
  \author[\instBNL]{D.~M.~Asner,} 
  \author[\instCincinnati]{H.~Atmacan,} 
  \author[\instHSE]{T.~Aushev,} 
  \author[\instTabuk]{R.~Ayad,} 
% ...

\emailAdd{souravpatra3012@gmail.com}

\abstract{We present a search for the charged lepton-flavor-violating decays $\yll$ and radiative charged lepton-flavour-violating decays $\ygll$ [$\ell,\ell^\prime = e, \mu, \tau$] using the 158 million $\y2$ sample collected by the Belle detector at the KEKB collider. This search uses $\y1$ mesons produced in $\ppy$ transitions. We do not find any significant signal, so we provide upper limits on the branching fractions at the 90\% confidence level.}

\begin{document}
 
\begin{flushright}
Belle Preprint 2022-02\\
KEK Preprint 2021-59\\
\end{flushright}
\maketitle
\flushbottom

\section{Introduction}
Observations of neutrino oscillations~\cite{neutrino} imply that the accidental lepton family symmetry in the standard model (SM) Lagrangian is broken. The minimal extension of the SM that can explain neutrino oscillations requires the presence of a right-handed neutrino. In such a framework, the conservation of individual lepton flavor is violated, and charged lepton-flavor-violating (CLFV) transitions can occur, mediated by $W^\pm$ bosons and massive neutrinos. However, the existence of such CLFV transitions would imply a minimal value of $\mt B\left(\mu^\pm \to e^\pm\gamma\right)\sim 10^{-54}$~\cite{intro1,intro2}. Several new physics models inspired by grand unified theories, such as supersymmetry and those predicting leptoquarks, typically enhance decay rates of CLFV transitions~\cite{gut1,gut2}.

The effective Lagrangian of new physics (NP) models can be expressed as the sum of a dipole term, four-fermionic interactions, and a gluonic interaction part. The Wilson coefficients of the NP operators can be determined via fits to measurements of phenomena those involve CLFV interactions~\cite{petrov}. Several classes of operator, such as vector, axial-vector, and tensor operators involved in four-fermionic interactions, allow CLFV transitions. Precise measurement of two-body vector meson CLFV decays allows one to probe the vector and tensor operators effectively.

A few results have previously been published related to two-body CLFV $\y{n}\rightarrow\ell\ell^\prime$ [$\ell,\ell^\prime = e, \mu, \tau$ and $n=1,2,3$] decays~\cite{cleo, babar}. Only the $\ymt$ decay has been studied, and no $\yem$ and $\yet$ results are available. CLFV $\y1$ decays can be studied with direct production or di-pion tagging of $\y2$ decays. Belle accumulated 6~$\fb$ of data at $\y1$ resonance, corresponding to 119 million $\y1$ events. However, it is difficult to judiciously trigger the two-charged-particle final state of these events: the sample is subject to extensive backgrounds, predominantly from QED processes. Belle also accumulated 25~$\fb$ of data at $\y2$ resonance, corresponding to 28 million $\y1$ produced in $\ppy$ decays. The four-particle final state of $\ppy$ allows for a more efficient trigger and for the suppression of the QED background. Therefore, we search for the $\yll$ decays using the $\y2$ data sample.

Radiative lepton-flavor-violating (RLFV) transitions allow one to probe the operators which are not easily accessible in the two-body decays~\cite{petrov}. Using three-body vector meson RLFV decays, one can put constraints on the corresponding Wilson coefficients of axial-vector, scalar, and pseudoscalar operators. Thus, RLFV studies of $\y{n}$ [$n=1,2,3$] provide complementary access to the NP parameters. Currently, there are no existing results available for the $\y{n}\rightarrow\gamma\ell^\pm\ell^{\prime\mp}$ decays. We perform the first search for RLFV in $\ygll$ decays using the $\y2$ data sample.

\section{Belle experiment}
The world's largest $\y2$ sample, corresponding to 158 million $\y2$ events, was collected with the Belle detector at the KEKB asymmetric-energy $e^+e^-$ collider~\cite{kekb} operating at a center-of-mass of energy  ($\sqrt{s}$) of 10.02 GeV. We study the $e^+e^-\rightarrow q\bar{q}~(q=u,d,s,c)$ background using the 80 $\fb$ data sample collected at 10.52 GeV. 

The Belle detector is a large-solid-angle spectrometer, which includes a silicon vertex detector (SVD), a 50-layer central drift chamber (CDC), an array of aerogel threshold Cherenkov counters (ACC), time-of-flight scintillation counters (TOF), and an electromagnetic calorimeter (ECL) comprised of 8736 CsI(Tl) crystals located inside a superconducting solenoid coil that provides a 1.5T magnetic field. An iron flux return located outside the coil is instrumented to detect $K^0_{\rm L}$ mesons and identify muons (KLM). The detector is described in detail elsewhere~\cite{detector}.

\section{Event selection}
We use the EVTGEN package~\cite{evtgen}, with QED final-state radiation simulated by PHOTOS~\cite{photos}, for the generation of Monte Carlo (MC) simulation events. We generate the signal events for two-body CLFV modes using both the vector to two leptons decay model, VLL and the phase space decay model, PHSP. The reconstruction efficiencies for the MC signal events generated with the VLL model are smaller (by approximately 8\%) than for the PHSP model. We will use the MC signal events generated with the VLL decay model for two-body CLFV decays to quote the most conservative upper limits. We use the PHSP model to generate the signal events for RLFV modes. We are using TAUOLA~\cite{tauola} or PYTHIA~\cite{pythia} for generating signal events for the subsequent decays of $\tau$ leptons. A GEANT3-based~\cite{geant} MC simulation is used to model the response of the detector. Thus, dedicated MC samples are generated for different signal modes to determine signal efficiencies and define selection criteria. Background studies and the optimization of those criteria are performed using an MC simulated sample of $\y2$ events with a size corresponding to the integrated luminosity. Dominant backgrounds arise from $\ppy$ decays with $\Yll~[\ell=e,\mu,\tau]$. The MC samples for these decays, corresponding to about 20 times of data sample sizes, are used to study backgrounds. For two-body CLFV searches, $\y1$ candidates are reconstructed in the $e^\pm\mu^\mp$, $\mu^\pm\tau^\mp$, and $e^\pm\tau^\mp$ final states. We reconstruct $\tau$ in $\te$, $\tP$ for the $\ymt$ decay and $\tm$, $\tP$ for the $\yet$ decay, comprising 28\% of $\tau$ decays. To avoid potential background from $\ymm$ ($\yee$) decays, we do not consider $\tm$ ($\te$) decays for the $\ymt$ ($\yet$) mode. Similarly, for RLFV decays we reconstruct $\y1$ in $\gamma e^\pm\mu^\mp$, $\gamma \mu^\pm\tau^\mp$, and $\gamma e^\pm\tau^\mp$ final states, where the $\tau$ is identified in $\te$ ($\tm$) decay for the $\ygmt$ ($\yget$) study. Also, we reconstruct $\yee$ and $\ymm$ decays, which are used to validate and calibrate the analysis. To validate the recoil $\y1$ sample along with muon and electron identifications, we measure the branching fractions of $\yee$, and $\ymm$ decays. As taus are mostly reconstructed in the leptonic decays, validating lepton identification with the high momentum leptons from $\y1$ is also important for $\ylt$ and $\yglt$ decays. This analysis follows a blind analysis procedure.

Charged tracks are required to originate from the vicinity of the interaction point (IP). The distance of the closest approach to the IP is required to be within 3.5~cm along the beam direction and within 1.5~cm in the transverse plane. The combined information from the CDC, TOF, and ACC is used to identify charged pions based on the pion likelihood ratio, ${\cal L}_\pi = {\cal P}_\pi/(\mt P_\pi+\mt P_K)$, where $\mt P_\pi$ and $\mt P_K$ are likelihood values for the pion and kaon hypotheses, respectively~\cite{pid}. Pions are required to have ${\cal L}_\pi > 0.6$, which has an identification efficiency of 94\%. Muon candidates are identified using a likelihood ratio, ${\cal L}_\mu$, which is based on the difference between the range of the track calculated from the particle momentum and that measured in the KLM. This ratio includes the value of $\chi^2$ formed from the KLM hit locations with respect to the extrapolated track. The muon identification efficiency for the applied selection, ${\cal L}_\mu$ > 0.95, is 89\%, with a pion misidentification probability of 1.4\%~\cite{muid}. Identification of electrons uses an analogous likelihood ratio, ${\cal L}_e$, based on specific ionization from the CDC, the ratio of the energy deposited in the ECL to the momentum measured by the CDC and SVD combined, the shower shape in the ECL, hit information from the ACC, and matching between the position of the charged track and the ECL cluster. The electron identification efficiency for the applied selection, ${\cal L}_e$ > 0.6, is 95\%, with a pion misidentification probability 0.3\%~\cite{eid}. To recover the bremsstrahlung energy loss for electrons and positrons, we include the energy from the photon(s) within 50~mrad of each of the $e^\pm$ tracks, which improves the efficiency for true signal events by 2.7\%. 

\begin{figure}[tbp]
\centering
\includegraphics[width=.65\textwidth]{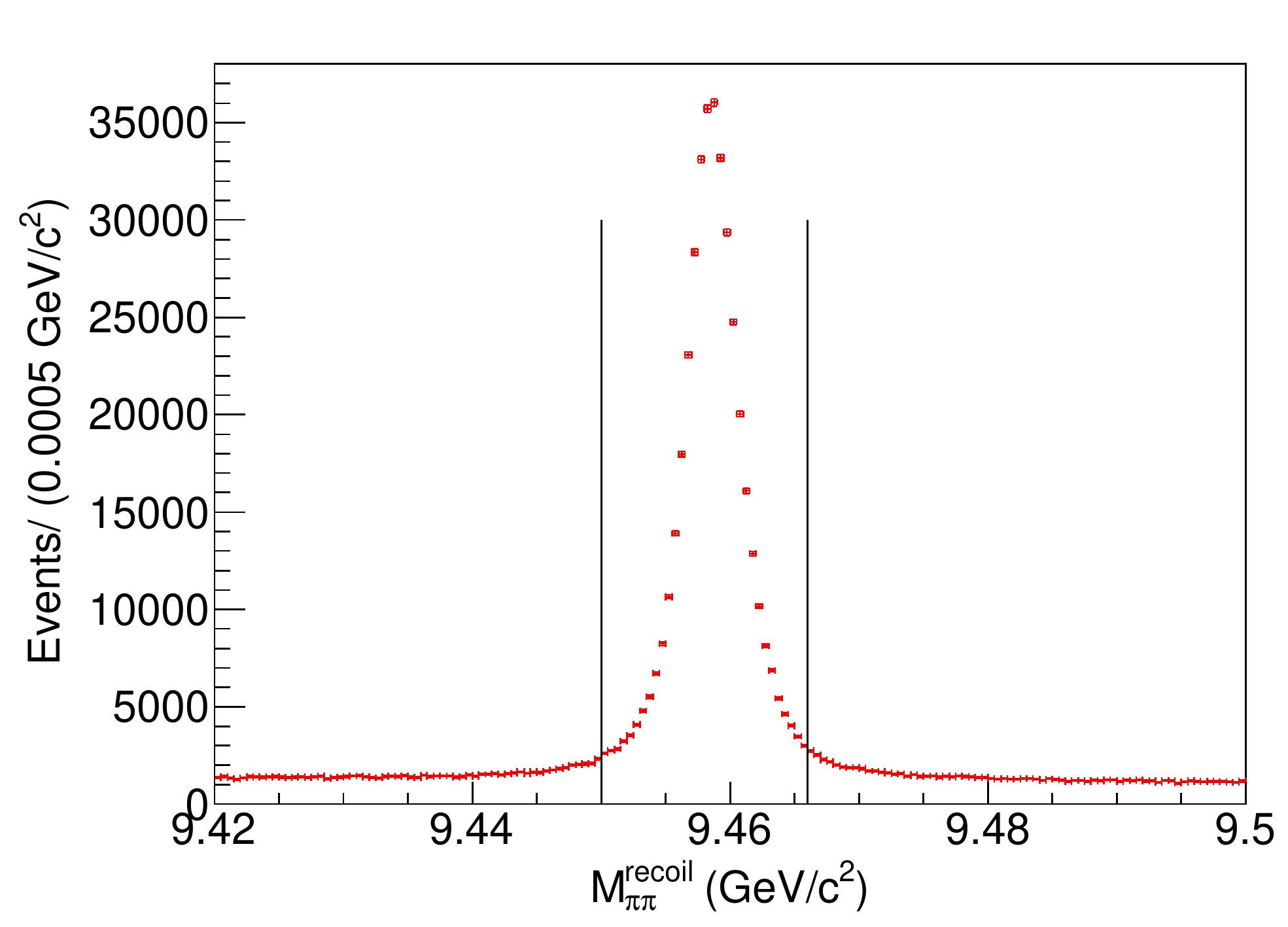}
\caption{\label{fig:mrecl}$\mrpp$ distribution for $\y2$ data. Events within the two vertical lines are selected.} 
\end{figure}

For $\ylt$ decay, most of the background comes from $\ytt$ and $\Yll$ decays. To suppress events coming from $\y2$ decays, other than $\ppy$, we define the recoil mass of two pions as:

\begin{equation}
\mrpp = \sqrt{(\E{total}-\E{\pi\pi})^2 - \p{\pi\pi}^2}
\label{eq:mrpp}
\end{equation}
Where $\E{total}$, $\E{\pi\pi}$, and $\p{\pi\pi}$ are the total energy of the colliding $e^+e^-$ beams, the energy of the two pions from $\y2$ and the magnitude of the 3-momentum of the pion pair, respectively in the center-of-mass (CM) frame. The distribution of $\mrpp$ is shown in Fig.~\ref{fig:mrecl}. The $\mrpp$ distribution peaks at the $\y1$ mass for signal events, while it is flat for the combinatorial background. We consider the events within $9.450 < \mrpp < 9.466$ $\gev$, corresponding to a $\pm3\sigma$ region around the nominal $\y1$ mass~\cite{pdg}. 
To suppress the background from $e^+e^-\gamma$ and $\mu^+\mu^-\gamma$, we remove the events with the cosine of the angle between the two pions in the $\y2$ rest frame ($\cospp$) greater than $0.5$. We define the visible tau momentum ($\pvt$) as the sum of the momentum carried by the daughter charged track(s) of $\tau$ in the lab frame. We select the $\tau$ candidates with $\pvt$ > 0.3 $\Gev$. Furthermore, $\tP$ is reconstructed with the invariant mass of the three-pion lower than 1.8 $\gev$ and energy in the lab frame greater than 2.6~GeV. These $\tau$ selections are wide enough to account for the missing momentum from neutrinos.
Also, we fit the three-pion vertex for the $\tP$ decay and events with fitted $\chi^2$ < 15 are selected to reduce combinatorial backgrounds. We count the number of tracks identified as muons or electrons with energy in the lab frame greater than 1~GeV as prompt muons ($\Nm$) and prompt electrons ($\Ne$), respectively. For $\ymt$ decays, in order to reject the background coming from the $\ymm$ decay, we select the events with $\Nm=1$ and $\Ne\le1$ ($\Ne=0$) for the $\te$ ($\tP$) reconstruction mode. Similarly, for $\yet$ decays, we select the events with $\Ne=1$ and $\Nm\le1$ ($\Nm=0$) for $\tm$ ($\tP$) reconstruction mode. We suppress a large number of $\ymm$ and $\yee$ backgrounds by the selections of prompt leptons.

\begin{figure}[tbp]
\centering
\includegraphics[width=.65\textwidth]{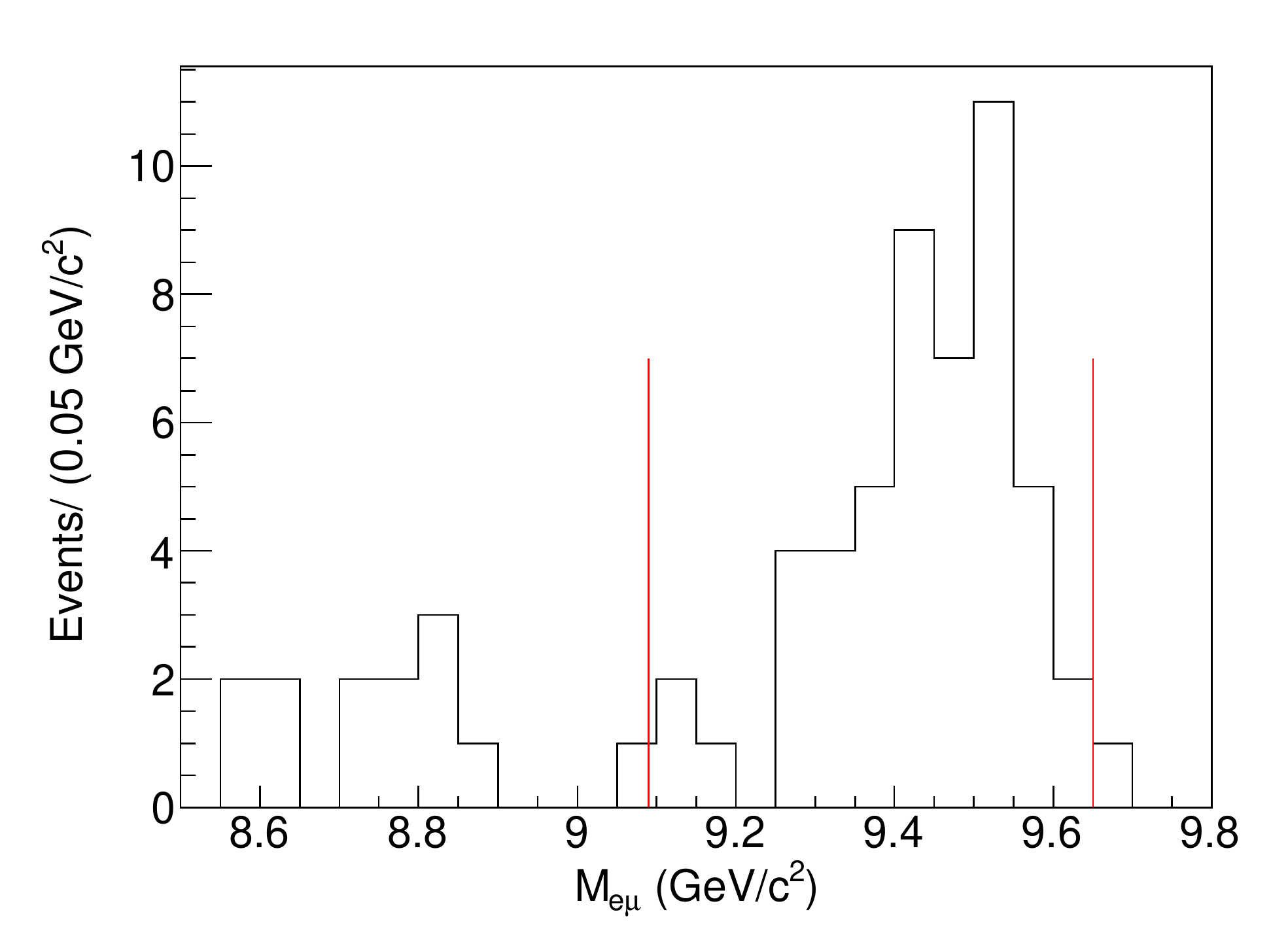}
\caption{\label{fig:my1}$\m{e\mu}$ distribution of $\pi^+\pi^-$ recoil sample in $\y2$ data. Events within two perpendicular lines are selected for $\yem$ decay.}
\end{figure}

For the $\yem$ study, the distribution of lepton pair invariant mass ($\m{e\mu}$) for $\pi^+\pi^-$ recoil sample in $\y2$ data is shown in Fig.~\ref{fig:my1}. We select the events with $\m{e\mu}$ within 9.09 to 9.65 $\gev$ by selecting a $\pm3\sigma$ region around the mean position and the $\y1$ momentum in the lab frame ($\p{e\mu}$) less than 4.4 $\Gev$ to reduce the $e^+e^-\rightarrow q\bar{q}$ events.
  
After applying all the selections, we find 3\% of events with multiple $\y2$ candidates for $\yem$ decay, and 4\% for both of $\ymt$ and $\yet$ decays. We performed a vertex fit with the reconstructed charged tracks of $\y2$, and the fitted $\chi^2$ value has been used to select the best candidate among the multiple $\y2$ candidates. Best candidate selection efficiencies for $\yem$, $\ymt$, and $\yet$ decays are 97\%, 89\%, and 91\%, respectively.

For RLFV decays, there is an extra photon in the final state. Therefore, we include all the selections which are used for the pion and lepton in the corresponding two-body CLFV decay previously discussed. For $\ygem$ decay, $\m{e\mu}$ and $\p{e\mu}$ of $\yem$ decay will be replaced by  $\m{\gamma e\mu}$ and $\p{\gamma e\mu}$, respectively. In addition to the above selections to the corresponding non-radiative mode, we select photons with energy in the lab frame greater than 200 MeV to remove soft photons and beam backgrounds. The photons used in the bremsstrahlung recovery are not considered in reconstructing the radiative candidates. Inside the signal search window, we find 3\%, 8\%, and 7\% multiple $\y2$ candidates for $\ygem$, $\ygmt$, and $\yget$ decays, respectively. Multiplicity due to misreconstructed charged particles is handled using a procedure similar to two-body CLFV decays. Multiplicity occurring from the multiple photon candidates is removed by selecting the event randomly. Best candidate selection efficiencies for $\ygem$, $\ygmt$, and $\yget$ are 93\%, 84\%, and 87\%, respectively.

\section{Study of $\yee,\mu^+\mu^-$ decays as calibration modes}
\label{sec:cntl_sample}
To study the calibration modes, we select events with lepton pair invariant mass ($\m{\ell\ell}$) within 9.09 to 9.65 $\gev$ and momentum of the reconstructed $\y1$ in the lab frame ($\p{\ell\ell}$) less than 4.4 $\Gev$. To extract the signal for $\Yll$ decays, we perform an unbinned maximum likelihood (UML) fits to $\dm = \m{\pi\pi\ell\ell} -\m{\ell\ell}$, where $\ell = e, \mu$. The signal probability density function (PDF) used is a  sum of two Gaussians sharing a common mean. Backgrounds from all the sources are flat in the $\dm$ window and small compared to the signal yields. We fit the background with a first-order Chebyshev polynomial. To account for any resolution difference between data and MC, the mean and the width parameter of the primary Gaussian ($\sigma_1$) are floated in the fit, and the width of the secondary Gaussian is set to $\sigma_2 = k\times\sigma_1$, with the factor $k$ fixed from MC.

\begin{figure}[tbp]
\centering
\includegraphics[width=.48\textwidth]{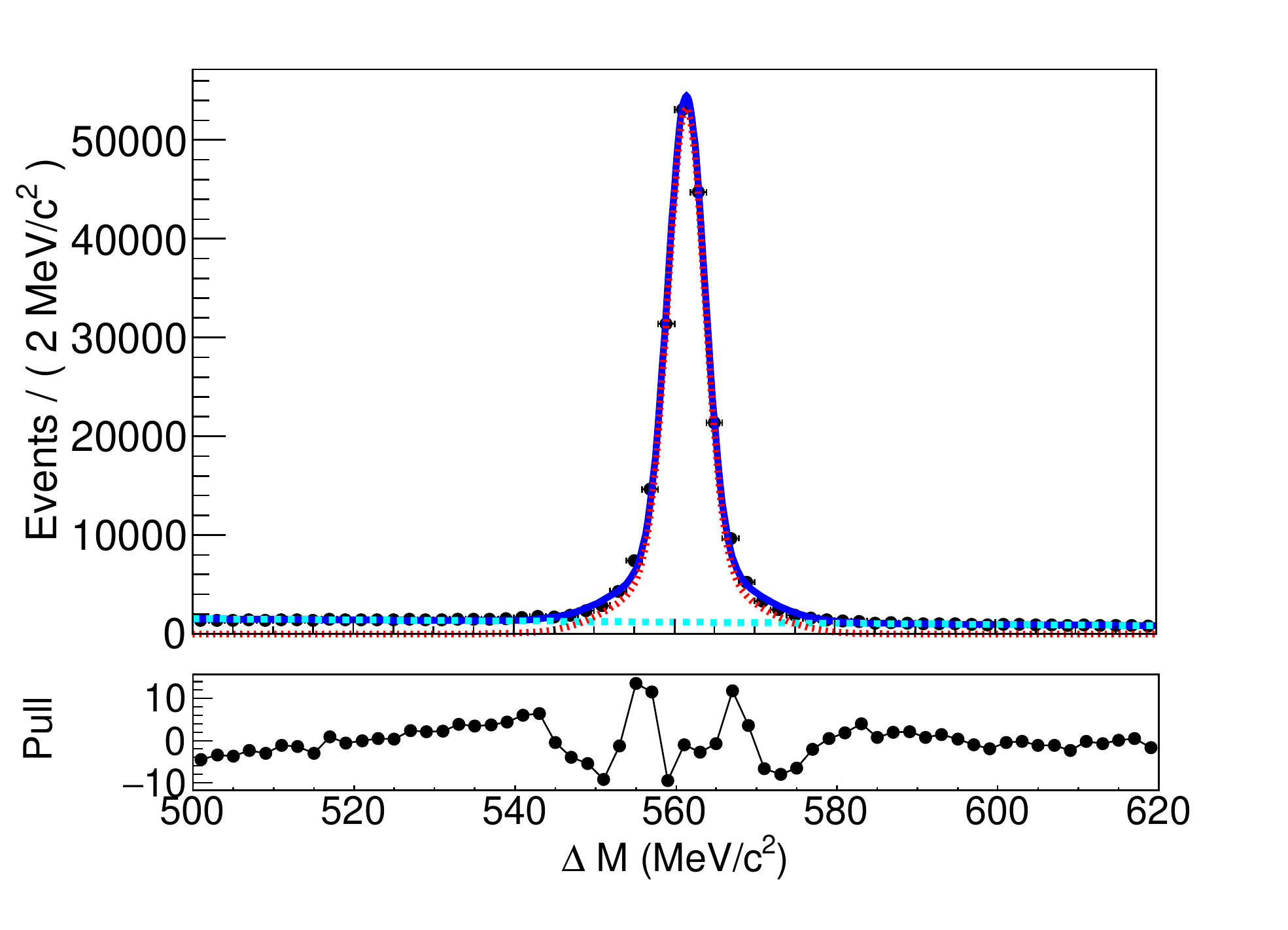}
\hfill
\includegraphics[width=.48\textwidth]{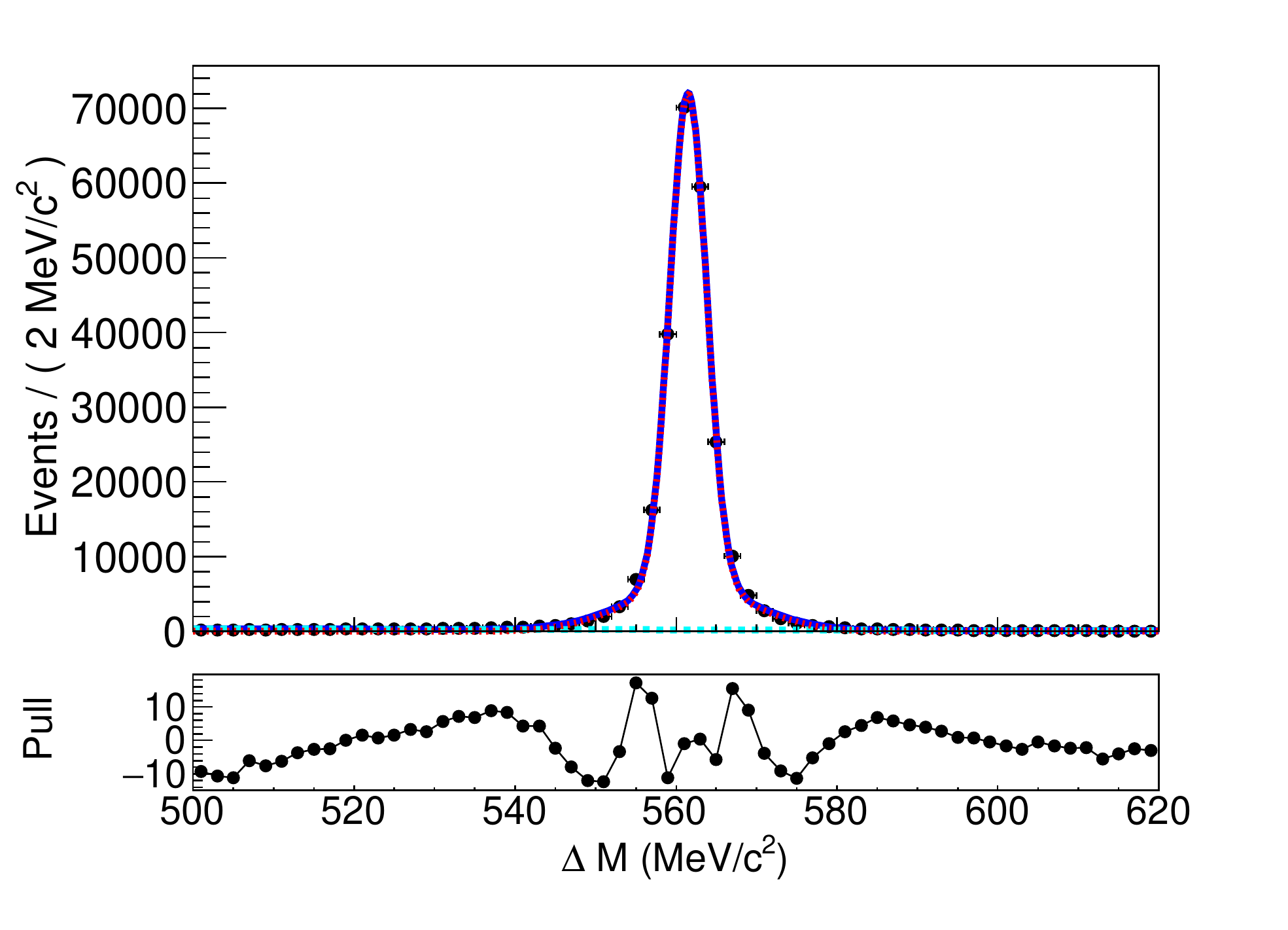}
\caption{\label{fit:ll}$\dm$ fit for $ee$ events (left) and $\mu\mu$ events (right). The dotted red curves represent the signal PDFs and the dashed cyan lines represent the background PDFs. The solid blue curves represent the overall fit to data.}
\end{figure}

Expected signal efficiencies for $ee$ and $\mu\mu$ are estimated to be 28.3\% and 35.6\%, respectively. Fig.~\ref{fit:ll} shows the fits to $\y2$ data. The signal yields obtained for the $ee$ and $\mu\mu$ final states are $191353 \pm467$ and $246255 \pm504$ events, respectively. The data-MC differences of widths for the $e^\pm e^\mp$ and $\mu^\pm\mu^\mp$ final states are estimated to be 12\% and 16\%, respectively. One can calculate the branching fractions using the following relation:

\begin{equation}
\mt{B}[\Yll] = \frac{\N{sig}}{N_{\y2}\times\mt{B}[\ppy]\times\epsilon}
\label{eq:bf}
\end{equation}
where, $N_{\y2}$, $\N{sig}$ and $\epsilon$ are the number of $\y2$ produced in $e^+e^-$ collision, signal yield in data and the effective signal efficiency (after implementing all the systematic corrections) respectively. Using Equation~\eqref{eq:bf}, the calculated branching fractions including only statistical uncertainties for $\yee$ and $\ymm$ are $(2.40\pm0.01)\times\e2$ and $(2.46\pm0.01)\times\e2$, respectively. These are consistent with the world average values~\cite{pdg}. These results are discussed further in Section~\ref{result} after including the systematic uncertainty. 

\section{Signal extraction for two-body CLFV decays}
\subsection{$\yem$ decay}
We extract the signal yield from a UML fit to the $\dm$ variable. $\dm$ should peak at the nominal mass difference between $\y2$ and $\y1$, approximately 560 $\mev$~\cite{pdg}. A sum of two Gaussians sharing a common mean has been used as the signal PDF. To estimate the peaking background, the shape of the peaking background is considered to have the same shape as the signal PDF. The $q\bar{q}$ backgrounds are flat, and they are modeled with a first-order Chebyshev polynomial. The width of the signal PDF in the data is fixed at the MC width, corrected by the average of the data-MC difference for the $\mu\mu$ and $ee$ samples.

\begin{figure}[tbp]
\centering
\includegraphics[width=.65\textwidth]{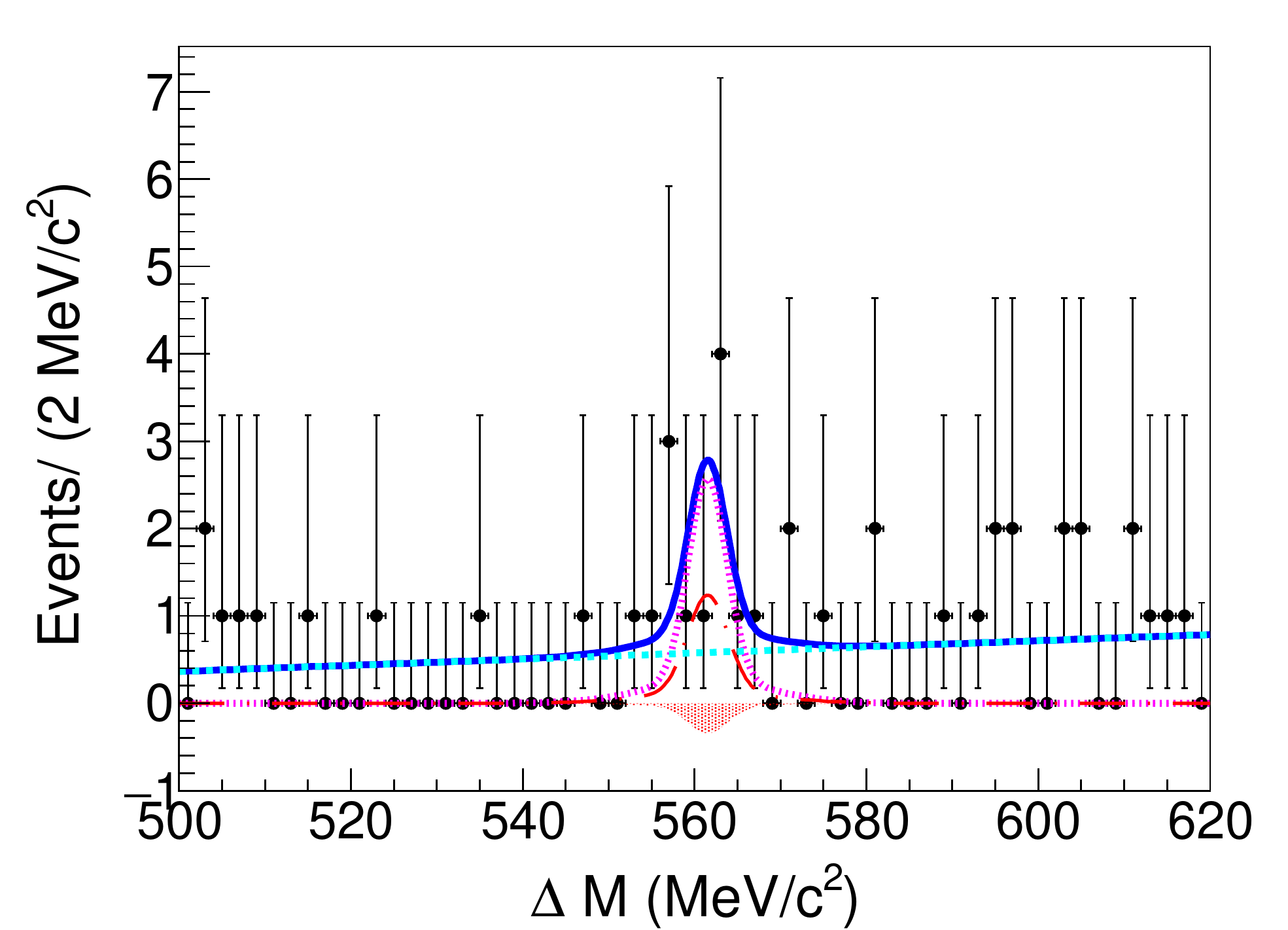}
\caption{\label{fit:em}$\dm$ fit to $\y2$ data for $\yem$ decay. The fitted signal PDF is represented by the filled red region, the dashed cyan line represents the flat background and the dotted magenta curve is the peaking background from lepton misidentification. The solid blue curve represents the overall fit to data. The long-dashed red curve represents the signal PDF corresponding to 5 hypothetical signal events.}
\end{figure} 

A few $\ymm$ ($\yee$) events mimic our signal when a $\mu^\pm$ ($e^\pm$) is misidentified as an $e^\pm$ ($\mu^\pm$). The amount of $\ytt$ background is estimated to be negligible. The number of $\ymm$ background events is estimated to be $3.5\pm0.4$ using a large MC sample. Such backgrounds are difficult to remove completely. To estimate the background from muon to electron misidentification in the data, we derive a correction factor for electron misidentification efficiency of muons using an $e^+e^-\rightarrow\mu^+\mu^-$ sample collected at $\sqrt{s}=10.52$~GeV with a tag-and-probe method. The data to MC correction factor for electron misidentification efficiency is estimated to be $2.5\pm0.5$, which leads to an estimation of this peaking background yield of $8.8\pm2.0$. The background from electron to muon misidentification is expected to be consistent with zero ($0.1\pm0.1$ events). Fitted distribution of $\y2$ data is shown in Fig.~\ref{fit:em}.  To consider the peaking background, we include a fixed PDF of 8.8 events in the data fit (dotted magenta line), and uncertainty (2.0) will be added in systematic uncertainty. The estimated signal efficiency for the $\yem$ mode is 32.5\%. We finally obtain a yield of  $-1.3\pm3.7$ signal events for the $\yem$ decay. 

\subsection{$\ylt$ decay}
For $\mu\tau$ and $e\tau$ decays of $\y1$, we extract the signal from an UML fit to the recoil mass of $\pi\pi\ell$ ($\mrppl$), where $\ell=\mu,e$. $\mrppl$ can be defined by replacing $\pi\pi$ with $\pi\pi\ell$ in Eq.~\eqref{eq:mrpp}. As $\mrppl$ is calculated from all the particles from the $\y2$ except the $\tau$, $\mrppl$ should peak at the nominal $\tau$ mass (around 1.78~$\gev$)~\cite{pdg}. Signal events of $\mu\tau$ and $e\tau$ decays are modeled with a sum of one Gaussian and one bifurcated Gaussian, sharing a common mean.
\begin{figure}[tbp]
\centering
\includegraphics[width=.62\textwidth]{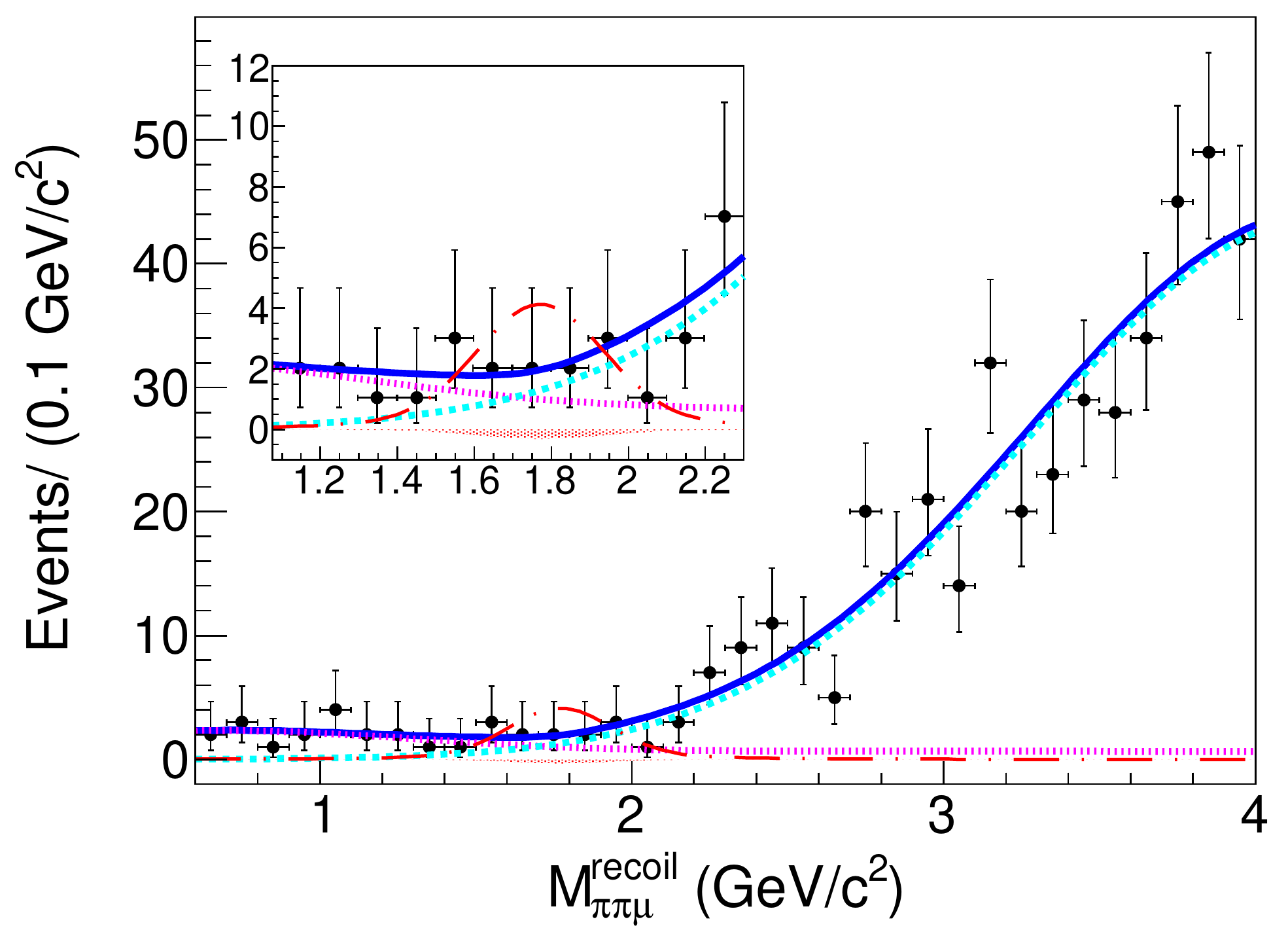}
\hfill
\includegraphics[width=.62\textwidth]{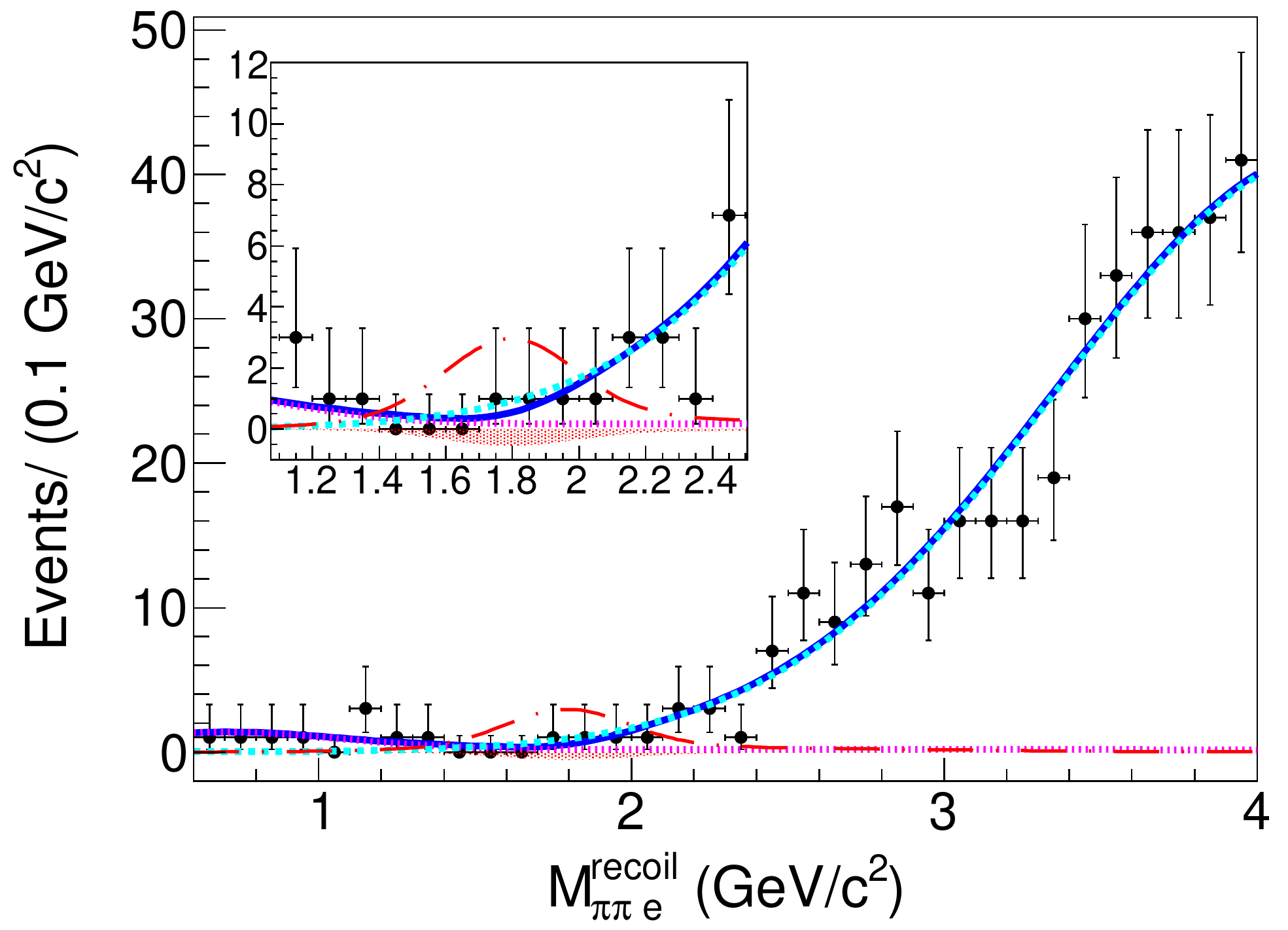}
\caption{\label{fit:lt}$\mrppl$ fit to $\y2$ data for the $\ymt$ decay (top) and the $\yet$ decay (bottom). The fitted signal PDFs are represented by the filled red regions. The dotted magenta line represents the contribution from $\mu\mu$ (or $ee$) background and the dashed cyan line represents the $\tau\tau$ background. The solid blue curves represent the overall fit to data. The long-dashed red curves represent the signal PDFs corresponding to 20 hypothetical signal events. To make the small signal yield in data visible, we add zoomed inset of the signal region.}
\end{figure}
We obtain a difference of 7\% (27\%) for the resolution between data and MC for $\mrppm$ ($\mrppe$) using the data-MC difference for the $\ymm$ ($\yee$) mode. The width of the $\ymt$ ($\yet$) signal PDF in the data is fixed from the MC width corrected by the data-MC difference for the $\mrppm$ ($\mrppe$) parameter.  

For $\ymt$ decays, the main backgrounds come from $\ytt$ and $\ymm$ decays. For the $\tau\tau$ background, a charged lepton or a pion from one of the tau decays is used as the signal muon. In the $\mrppm$ distribution, missing neutrino energy from the misidentified $\tau$ shifts such events away from the actual $\tau$ mass. Thus, the $\tau\tau$ background increases exponentially, starting near the nominal $\tau$ mass value. We model the $\tau\tau$ background using the following exponential threshold PDF starting near $M_{\rm th}$ (in $\gev$),

\begin{equation} 
\mt F(\mrppl;A,B,M_{\rm th})=\exp{[A(\mrppl-M_{\rm th})+B(\mrppl-M_{\rm th})^2]}
\label{eq:tau_pdf}
\end{equation}
{where $A$ and $B$} are the two slope parameters of the $\tau\tau$ background PDF. We try other fitting models and find the current model describes the background the best. To get the proper shape of the $\tau\tau$ background PDF, we use a large $\ytt$ MC sample. In the data fit, $M_{\rm th}$ and $B$ are fixed from the MC background, and $A$ is allowed to float. Fig.~\ref{fit:lt} shows the fitted distributions of data for $\ylt$ decays. The background from $\ymm$ is obtained from a large MC sample. The expected number of $\mu\mu$ backgrounds is less than the number of $\tau\tau$ backgrounds, but it widely populates around the signal region: it peaks at the lower mass value ($<1~\gev$) and has a broad tail. The PDF is presented by a sum of one bifurcated Gaussian and one threshold function starting from 0 $\gev$ using a large $\ymm$ sample. To fit the data, we float the yield of the $\mu\mu$ background fixing the shape of the Gaussian of $\mu\mu$ background from MC corrected by the data-MC difference for the $\mrppm$ parameter.

For $\yet$ decays, potential backgrounds arise from $\ytt$ and $\yee$ decays. These backgrounds are handled using a procedure similar to that used for the $\tau\tau$ and $\mu\mu$ backgrounds to $\ymt$ decays.

The expected peaking backgrounds in the $\y2$ data for $\ymt$ and $\yet$ decays are estimated from MC to be $0.7\pm4.1$ and $4.6\pm5.4$, respectively. As no significant peaking background is found in the $\y2$ decay MC sample, we do not include a peaking background component in the fit. Considering both the $\tau$ reconstruction modes, the effective signal efficiency for $\ymt$ ($\yet$) decay is 8.8\% (7.1\%). In $\y2$ data, we find the yield of $\ymt$ and $\yet$ signals to be $-1.5\pm4.3$ and $-3.5\pm2.7$, respectively. Hence, there is no evidence for $\ylt$ transitions.

\section{Signal extraction for RLFV decays}
\subsection{$\ygem$ decay}
\begin{figure}[tbp]
\centering
\includegraphics[width=.65\textwidth]{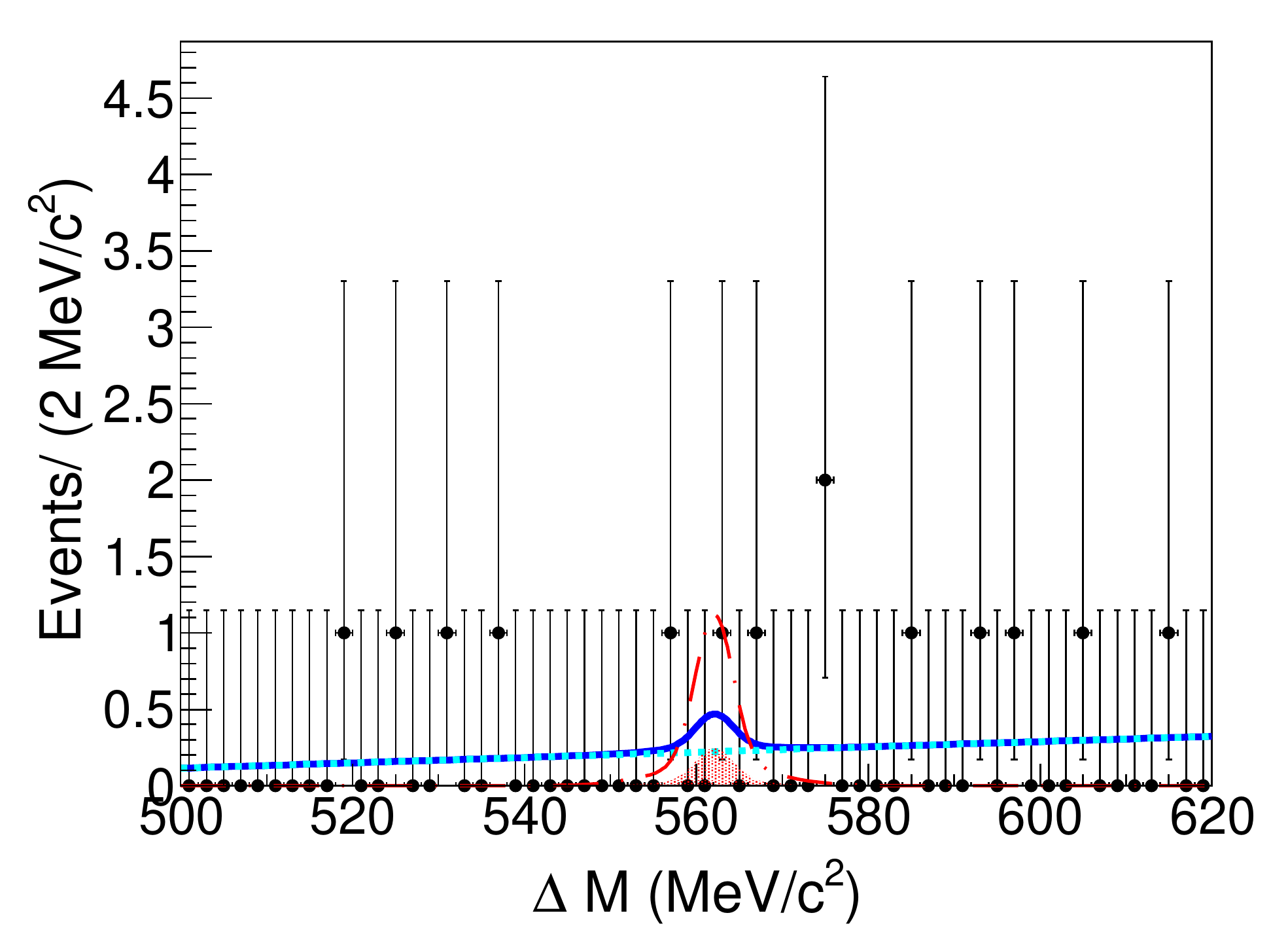}
\caption{\label{fit:gem}$\dm$ fit to $\y2$ data for the $\ygem$ decay. The fitted signal PDF is represented by the filled red region and the dashed cyan line represents the background. The solid blue curve represents the overall fit to data. The long-dashed red curve represents the signal PDF corresponding to 5 hypothetical signal events.}
\end{figure}

Our RLFV signal extraction procedure is very similar to that used for the corresponding non-radiative transition. We perform an UML fit to the mass difference $\dm = \m{\pi\pi\gamma e\mu} -\m{\gamma e\mu}$. The signal PDF used is a sum of two Gaussians sharing a common mean. To estimate the peaking background from leptonic decays of the $\y1$, we use the same shape as the signal PDF as the shape of the background PDF and a large $\Yll$ MC sample is used to have a more precise estimation. Other backgrounds are flat on the $\dm$ window and modeled with a first-order Chebyshev polynomial. Fig.~\ref{fit:gem} shows the $\dm$ fit for $\y2$ data. To fit the data, we fix the width of the signal PDF from the MC signal width corrected by the average of data-MC difference for $\yee$ and $\ymm$ modes. The yield of peaking background is estimated to be $0.1\pm0.1$. 
The signal efficiency for $\gamma e\mu$ decay is 24.6\%. From the $\y2$ data fit, the signal yield for the $\ygem$ decay is estimated to be $0.8\pm1.5$.

\subsection{$\yglt$ decay}
To extract the signal for $\gamma\mu\tau$ and $\gamma e\tau$ decays, we define the recoil mass of $\pi\pi\gamma\ell$ ($\mrppgl$) using Eq.~\eqref{eq:mrpp}. We perform a UML fit to $\mrppgl$ to extract the signal yield and estimate efficiency. For signal events, $\mrppgl$ should peak at the nominal $\tau$ mass. A Gaussian and a bifurcated Gaussian sum sharing a common mean is used to model the signal events for $\yglt$ decays. 
\begin{figure}[tbp]
\centering
\includegraphics[width=.65\textwidth]{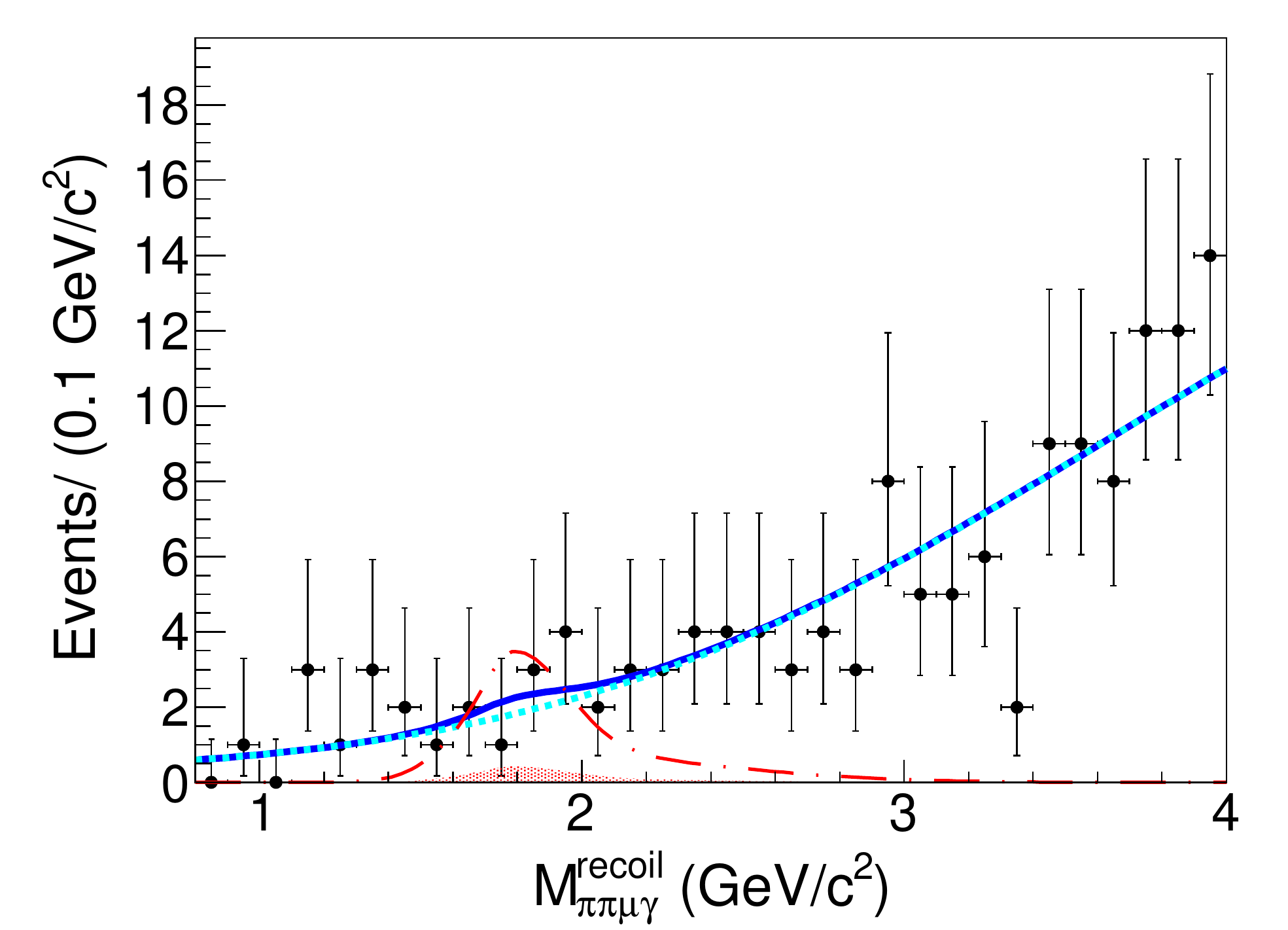}
\hfill
\includegraphics[width=.65\textwidth]{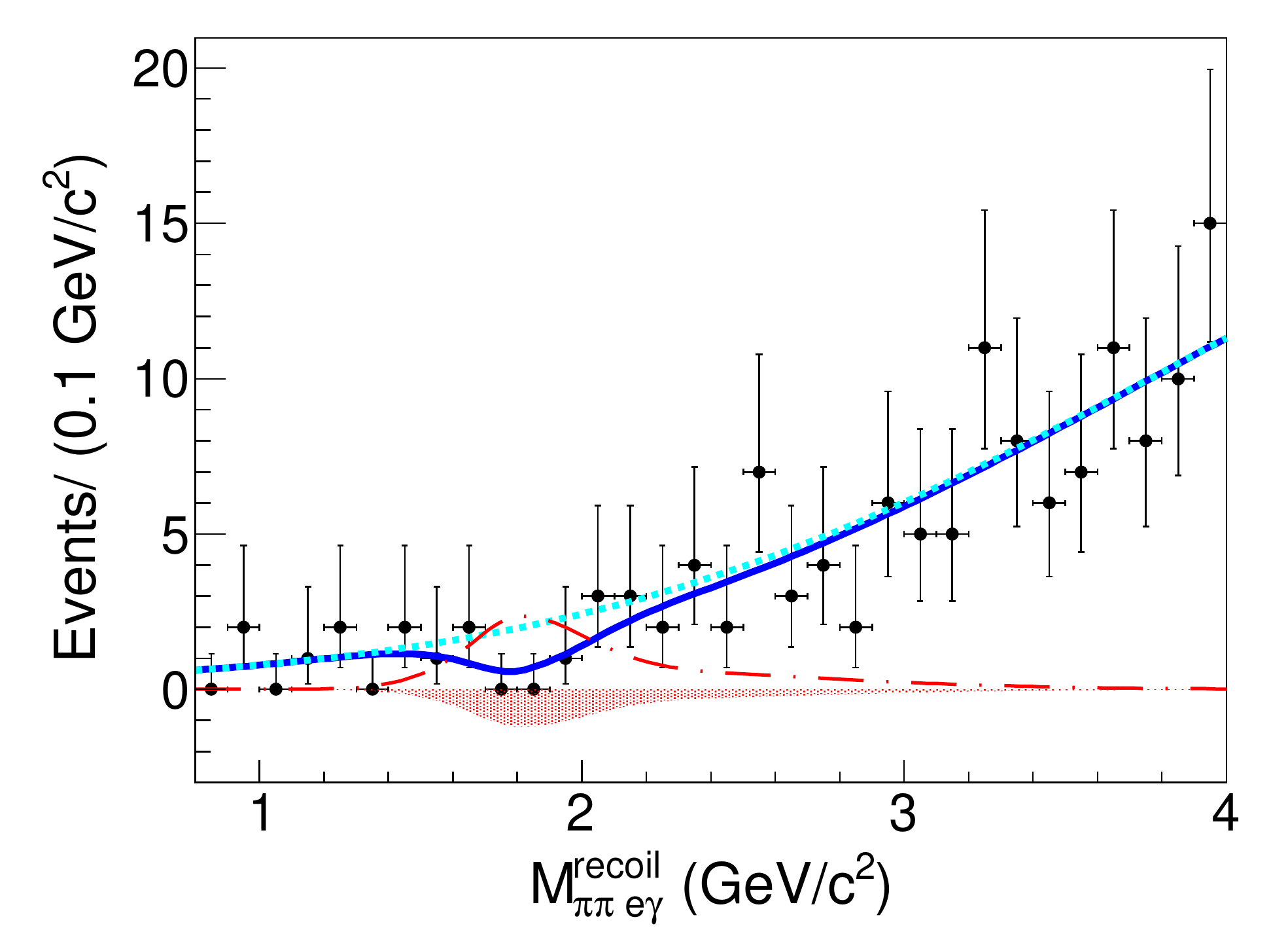}
\caption{\label{fit:glt}$\mrppgl$ fit to $\y2$ data for the $\ygmt$ decay (top) and the $\yget$ decay (bottom). The fitted signal PDFs are represented by the filled red regions and the dashed cyan lines represent the total background. The solid blue curves represent the overall fit to data. The long-dashed red curves represent the signal PDFs corresponding to 20 hypothetical signal events.}
\end{figure}  

For both of the $\yglt$ decays, the dominant background comes from $\tau\tau$ decays and hadronic decays of the $\y1$. The $\tau\tau$ background is treated using an approach similar to that for the $\ymt$ decay, with the background shape as described by the Eq.~\eqref{eq:tau_pdf}. Also, we find some background from the radiative hadronic decays of $\y1$. We fit hadronic and all other backgrounds with an exponential PDF. The expected yield of peaking background events for $\ygmt$ decay is estimated to be consistent with zero ($1.8\pm5.1$). Similarly, the expected yield for $\yget$ decay is estimated to be $-7.9\pm7.4$. To fit the data, we fix all the parameters of the background PDF (including the shape of the exponential PDF and fraction of $\tau\tau$ background) from MC except the $A$ parameter of the $\tau\tau$ PDF. The fractions of $\tau\tau$ PDF for $\ygmt$ and $\yget$ decays are 58\% and 86\%, respectively. The width of the signal PDF for $\ygmt$ ($\yget$) decay in the data is fixed from MC signal width corrected by the data-MC difference for the $\mrppm$ ($\mrppe$) parameter. In Fig.~\ref{fit:glt}, we show the $\mrppgl$ fits to $\y2$ data. The effective signal efficiency for $\gamma\mu\tau$ ($\gamma e\tau$) decay is 5.8\% (5.0\%). The fitted signal yield for $\ygmt$ decays in the $\y2$ data is estimated to be $2.1\pm5.9$. Similarly, the signal yield for $\yget$ decay is obtained to be $-9.5\pm6.3$. Hence, there is no evidence for $\yglt$ transitions. 

\section{Systematic uncertainty and correction}
We calculate the systematic uncertainty from various sources such as the number of $\y2$, track reconstruction, photon reconstruction, identification of pions from $\y2$, lepton identification, uncertainty in signal efficiency, secondary branching fraction, and the fitting model. 

The uncertainty on the number of $\y2$ events was determined from a study of hadronic decays to be 2.3\%~\cite{ny2s}. Reconstruction efficiency of charged particle tracks are studied using a partially reconstructed ${D^{* +}}\rightarrow D^0[K_S^0(\pi^+\pi^-)\pi^+\pi^-]\pi^+$ decay sample with $p_{\rm T}>200$ $\Mev$. Systematic uncertainty per track is estimated to be 0.35\%. Due to correlation, uncertainties in charged track finding are added linearly. The efficiency of photon reconstruction is estimated with radiative Bhabha events, and the associated uncertainty is 2.0\%~\cite{t2lg}.

Uncertainty from pion identification in $\ppy$ reconstruction may affect our results. In order to estimate it, we use the results of a dedicated study based on the $D^{* +}\rightarrow D^0(K^-\pi^+)\pi^+$ decay. A correction for the difference in efficiency (between data and signal MC) is obtained from the same source. This correction is used to correct the efficiency, and its uncertainty is included as the systematic uncertainty due to pion identification. For all the decays, the efficiency correction factor and systematic uncertainty from pion pair reconstruction are estimated to be 1.00 and 1.9\%, respectively. For the electron identification with ${\cal L}_e > 0.6$ and the muon identification with ${\cal L}_\mu > 0.95$, systematic uncertainty are calculated from the comparison between data and MC for $2\gamma\rightarrow ee/\mu\mu$ decays. We calculate an efficiency correction factor and systematic uncertainty for all of the electrons and muons using the same approach. For $\tP$ decay, the systematic uncertainty due to pion identification is estimated using the $D^{* +}$-based method described above. The efficiency correction factors associated with the leptons pair reconstructions for $e\mu$, $\mu\tau$, $e\tau$, $\gamma e\mu$, $\gamma\mu\tau$, and $\gamma e\tau$ decays are 0.99, 0.98, 0.97, 0.95, 0.94, and 0.97, respectively and corresponding systematic uncertainties are 1.9\%, 2.1\%, 2.3\%, 2.6\%, 2.8\%, and 2.5\%, respectively.

Due to the limited number of generated MC signal events, there is an uncertainty in the fitted number of signal events as well as in the signal efficiency ($\epsilon$), and the corresponding uncertainty is included in the systematic uncertainty.
 
\begin{table}[tbp]
  \begin{center}
    \begin{tabular}{l|cccccccc}
    \hline
    \hline
    \multirow{2}{*}{Source} & \multicolumn{8}{c}{Systematic uncertainty(\%)}                     \\ \cline{2-9}
      &$\mt S_{e\mu}$ &$\mt S_{\mu\tau}$ &$\mt S_{e\tau}$ &$\mt S_{\gamma e\mu}$ &$\mt S_{\gamma\mu\tau}$ &$\mt S_{\gamma e\tau}$ &$\mt S_{ee}$ &$\mt S_{\mu\mu}$ \\\hline
      Number of $\y2$ &2.3 &2.3 &2.3 &2.3 &2.3 &2.3 &2.3 &2.3 \\ 
      Track reconstruction &1.4 &1.5 &1.5 &1.4 &1.4 &1.4 &1.4 &1.4 \\ 
	  Photon reconstruction &- &- &- &2.0 &2.0 &2.0 &- &- \\ 
      Reconstruction of $\pi^+\pi^-$ from $\y2$ &1.9 &1.9 &1.9 &1.9 &1.9 &1.9 &1.9 &1.9 \\ 
      1st lepton identification &1.6 &1.1 &1.5 &1.7 &2.0 &1.7  &1.7 &1.2\\ 
      2nd lepton identification &1.1 &1.8 &1.7 &2.0 &1.9 &1.8 &1.7 &1.2\\ 
       MC statistics &0.2  &0.3 &0.6 &0.3 &0.4 &0.4 &0.4 &0.3\\ 
      Secondary branching fractions &1.5 &1.5 &1.5 &1.5 &1.5 &1.5 &1.5 &1.5\\ 
      Fitting model &0.1 &1.5 &0.9 &0.1 &0.8 &1.5 &0.1 &0.0\\ \hline
      Total &4.1 &4.5 &4.4 &4.9 &5.1 &5.1 &5.0 &4.4 \\ \hline \hline
    \end{tabular}
  \end{center}
  \caption{\label{tab:syst}Summary of the systematic uncertainties for the measurement of branching fractions of the $\yem$ ($\mt S_{e\mu}$), $\ymt$ ($\mt S_{\mu\tau}$), $\yet$ ($\mt S_{e\tau}$), $\ygem$ ($\mt S_{\gamma e\mu}$), $\ygmt$ ($\mt S_{\gamma\mu\tau}$), $\yget$ ($\mt S_{\gamma e\tau}$), $\yee$ ($\mt S_{ee}$), and $\ymm$ ($\mt S_{\mu\mu}$) decays.}
\end{table}
To obtain the final results, previously measured branching fractions of $\ppy$, $\tm$, $\te$, and $\tP$ are used~\cite{pdg}. The uncertainties in the world average secondary branching ratios are included as systematic uncertainties. 

We fix some parameters (such as the mean, width, and fractions of the two Gaussians) of the signal PDF while fitting the data. The associated systematic uncertainties are estimated by varying each of the fixed parameters by $\pm 1\sigma$ from their central values and repeating the fit. For $\yee$ and $\ymm$ decays, these PDF systematic uncertainties are estimated to be 0.08\% and 0.04\%, respectively. In the absence of significant signal events for the CLFV modes, we take the average value of the control modes (0.06\%) as the systematic uncertainty from the signal PDF for each of the CLFV modes. Similarly, the systematic uncertainty from the background PDF for $\ymt$, $\yet$, $\ygmt$, and $\yget$ are estimated to be 1.4\%, 0.8\%, 0.7\%, and 1.4\%, respectively. No parameters were fixed to estimate the background for $\yem$, $\ygem$, $\yee$, and $\ymm$ modes. Therefore, for the above decays, the systematic uncertainty from the background PDF is estimated as zero. Due to correlation, systematic uncertainties from signal PDF and background PDF are added linearly. Table~\ref{tab:syst} summarizes the systematic uncertainties from various sources for all the modes. Systematic uncertainties from the different sources are added in quadrature in order to get the total systematic uncertainty for a particular signal mode. The systematic uncertainty due to the uncertainty in the peaking background for $\yem$ ($8.8\pm2.0$) is directly included in the estimated upper limit of the branching fraction. For other modes, the effect of possible peaking background lowers the upper limit, and we do not consider it to report conservative upper limits.

\section{Results}
\label{result}
Using equation~\eqref{eq:bf}, the branching fractions are calculated as $\mt B[\yee] = (2.40\pm0.01({\rm stat})\pm0.12({\rm syst}))\times\e2$ and $\mt B[\ymm] = (2.46\pm0.01({\rm stat})\pm0.11({\rm syst}))\times\e2$ which agree within $\pm1\sigma$ with world average values~\cite{pdg}. All of the results for the branching fractions of CLFV modes are dominated by statistical uncertainty. In the absence of significant signal, we estimate the upper limits (UL) of the branching fractions with a frequentist approach~\cite{ul}. One can calculate the UL of branching fractions using the following relation:

\begin{equation}
\mt{B}[\yll]<\frac{\N{sig}^{\rm UL}}{N_{\y2}\times\mt{B}[\ppy]\times\epsilon}
\label{eq:ul}
\end{equation}
where $\N{sig}^{\rm UL}$ is the UL on the signal yield after including systematic uncertainty. We perform 5000 pseudo-experiments by generating the fixed background from the final PDF and varying the yield of the input signal within 1 to 20. We use the corresponding PDF that has been used to fit $\y2$ data for generating the data sets for pseudo-experiments. The fraction of pseudo-experiments with a fitted yield greater than the estimated signal yield in data has been taken as the confidence level (CL). Systematic uncertainties of the CLFV modes are included by smearing the yield of the pseudo-experiments within the fluctuations. For $\yem$ decay, the fitted signal yields of pseudo-experiments have been smeared within the corresponding uncertainty of peaking background to include the associated systematic uncertainty.

For $e\mu$, $\mu\tau$, and $e\tau$ decays, the central values of signal yields are obtained as $-1.3$, $-1.4$, and $-3.5$, respectively. The fraction of pseudo-experiments with any positive yield has been used to estimate the 90\% CL upper limits. Considering the number of $\y2$ as 157.8 million and $\mt B [\ppy]$ as $(17.85\pm0.26)$\%, we calculate the ULs of branching fractions ($\mt B^{\rm UL}$) by substituting $\N{sig}^{\rm UL}$ in Eq.~\eqref{eq:ul}. The estimated ULs for $\yem$, $\ymt$, and $\yet$ at 90\% CL are $3.9\times\e7$, $2.7\times\e6$, and $2.7\times\e6$ respectively. We summarize these results in Table~\ref{tab:rslt}.
\begin{table}[tbp]
\centering
\begin{tabular}{lp{0.9cm}cccc}
\hline \hline
Decay &$\epsilon$ (\%) &$N_{\rm sig}^{\rm fit}$ &$\N{sig}^{\rm UL}$ &$\mt B^{\rm UL}$ &PDG result\\
 \hline
 $\yem$ &32.5 &$-1.3\pm3.7$ &3.6 &$3.9\times\e7$ &$-$\\ 
 $\ymt$ &8.8 &$-1.5\pm4.3$ &6.8 &$2.7\times\e6$ &$6.0\times\e6$\\ 
 $\yet$ &7.1 &$-3.5\pm2.7$ &5.3 &$2.7\times\e6$ &$-$\\ 
 $\ygem$ &24.6 &$+0.8\pm1.5$ &2.9 &$4.2\times\e7$ &$-$\\ 
 $\ygmt$ &5.8 &$+2.1\pm5.9$ &10.0 &$6.1\times\e6$ &$-$\\ 
 $\yget$ &5.0 &$-9.5\pm6.3$ &9.1 &$6.5\times\e6$ &$-$\\ \hline \hline
\end{tabular}
\caption{\label{tab:rslt}Results of searches for CLFV in $\y1$ decays. Here, $N_{\rm sig}^{\rm fit}$ is the fitted signal yield. $N_{\rm sig}^{\rm UL}$ and $\mt B^{\rm UL}$ are, respectively, the upper limits of signal yield and branching fraction at 90\% CL.}
\end{table}

For $\gamma e\mu$ and $\gamma\mu\tau$ decays, the central values of signal yields are estimated to be 0.8 and 2.1, respectively. The estimated signal yield for $\yget$ decay is $-9.5$. The fraction of pseudo-experiments with a signal yield greater than 0.8 (2.1) has been taken as the CL for $\gamma e\mu$ ($\gamma \mu\tau$) decay. For $\gamma e\tau$ decay, we treat the fraction of pseudo-experiments with a positive signal yield as the CL. Table~\ref{tab:rslt} summarizes the $\N{sig}^{\rm UL}$ for RLFV modes at the 90\% CL. Using $\N{sig}^{\rm UL}$ in Eq.~\eqref{eq:ul}, the ULs of branching fractions for $\ygem$, $\ygmt$, and $\yget$ are estimated to be $4.2\times\e7$, $6.1\times\e6$, and $6.5\times\e6$, respectively.

\section{Summary}
In this paper, we report the searches for charged lepton-flavor-violation in $\yll$ decays and radiative lepton-flavor-violation in $\ygll$ decays conducted at the Belle experiment, where $\ell,\ell^\prime = e, \mu, \tau$. The searches are based on the 28 million $\pi^+\pi^-\y1$ decays produced from 25~$\fb$ of $e^+e^-$ collisions collected at the $\y2$ resonance. We study the sources of possible background using a large $\y2$ MC sample. To validate the signal extraction procedure we measure the branching fractions for $\yee$ and $\ymm$ modes and find $\mt B[\yee] = (2.40\pm0.01({\rm stat})\pm0.12({\rm syst}))\times\e2$ and $\mt B[\ymm] = (2.46\pm0.01({\rm stat})\pm0.11({\rm syst}))\times\e2$, respectively. In the absence of signal, we set upper limits on the branching fractions of the CLFV decays at the 90\% CL. The result for the $\ymt$ decay is 2.3 times more stringent than the previous result from the CLEO collaboration~\cite{cleo}, while the remaining modes are searched for the first time.

\acknowledgments
We thank the KEKB group for the excellent operation of the
accelerator; the KEK cryogenics group for the efficient
operation of the solenoid; and the KEK computer group, and the Pacific Northwest National
Laboratory (PNNL) Environmental Molecular Sciences Laboratory (EMSL)
computing group for strong computing support; and the National
Institute of Informatics, and Science Information NETwork 5 (SINET5) for
valuable network support.  We acknowledge support from
the Ministry of Education, Culture, Sports, Science, and
Technology (MEXT) of Japan, the Japan Society for the 
Promotion of Science (JSPS), and the Tau-Lepton Physics 
Research Center of Nagoya University; 
the Australian Research Council including grants
DP180102629, 
DP170102389, 
DP170102204, 
DE220100462, 
DP150103061, 
FT130100303; 
Austrian Federal Ministry of Education, Science and Research (FWF) and
FWF Austrian Science Fund No.~P~31361-N36;
the National Natural Science Foundation of China under Contracts
No.~11675166,  
No.~11705209;  
No.~11975076;  
No.~12135005;  
No.~12175041;  
No.~12161141008; 
Key Research Program of Frontier Sciences, Chinese Academy of Sciences (CAS), Grant No.~QYZDJ-SSW-SLH011; 
the Shanghai Science and Technology Committee (STCSM) under Grant No.~19ZR1403000; 
the Ministry of Education, Youth and Sports of the Czech
Republic under Contract No.~LTT17020;
the Czech Science Foundation Grant No. 22-18469S;
Horizon 2020 ERC Advanced Grant No.~884719 and ERC Starting Grant No.~947006 ``InterLeptons'' (European Union);
the Carl Zeiss Foundation, the Deutsche Forschungsgemeinschaft, the
Excellence Cluster Universe, and the VolkswagenStiftung;
the Department of Atomic Energy (Project Identification No. RTI 4002) and the Department of Science and Technology of India; 
the Istituto Nazionale di Fisica Nucleare of Italy; 
National Research Foundation (NRF) of Korea Grant
Nos.~2016R1\-D1A1B\-01010135, 2016R1\-D1A1B\-02012900, 2018R1\-A2B\-3003643,
2018R1\-A6A1A\-06024970, 2019K1\-A3A7A\-09033840,
2019R1\-I1A3A\-01058933, 2021R1\-A6A1A\-03043957,
2021R1\-F1A\-1060423, 2021R1\-F1A\-1064008;
Radiation Science Research Institute, Foreign Large-size Research Facility Application Supporting project, the Global Science Experimental Data Hub Center of the Korea Institute of Science and Technology Information and KREONET/GLORIAD;
the Polish Ministry of Science and Higher Education and 
the National Science Center;
the Ministry of Science and Higher Education of the Russian Federation, Agreement 14.W03.31.0026, 
and the HSE University Basic Research Program, Moscow; 
University of Tabuk research grants
S-1440-0321, S-0256-1438, and S-0280-1439 (Saudi Arabia);
the Slovenian Research Agency Grant Nos. J1-9124 and P1-0135;
Ikerbasque, Basque Foundation for Science, Spain;
the Swiss National Science Foundation; 
the Ministry of Education and the Ministry of Science and Technology of Taiwan;
and the United States Department of Energy and the National Science Foundation.



\begin{thebibliography}{99}
\bibitem{neutrino}
Y. Fukuda {\it et al.} (Super-Kamiokande Collaboration),
Phys. Rev. Lett. {\bf 81}, 1562 (1998); Q. R. Ahmad {\it et al.} (SNO Collaboration), Phys. Rev. Lett. {\bf 89}, 011301 (2002).

\bibitem{intro1}
M.~Raidal {\it et al.}, Eur. Phys. J. C {\bf 57}, 13 (2008).

\bibitem{intro2}
A.M.~Teixeira, J. Phys. Conf. Ser. {\bf 888}, 012029 (2016).
\bibitem{gut1}
H.~Georgi and S.L.~Glashow, Phys. Rev. Lett. {\bf 32}, 438 (1974).

\bibitem{gut2}
J.C.~Pati and A.~Salam, Phys. Rev. D {\bf 10}, 275 (1974).

\bibitem{petrov}
D.E.~Hazard and A.A.~Petrov, Phys. Rev. D {\bf 94}, 074023 (2016).

\bibitem{cleo}
W.~Love {\it et al.} (CLEO Collaboration), Phys. Rev. Lett. {\bf 101}, 201601 (2008).

\bibitem{babar}
J.P.~Lees {\it et al.} (BaBar Collaboration), Phys. Rev. Lett. {\bf 104}, 151802 (2010).

\bibitem{kekb}
S.~Kurokawa, E.~Kikutani, Nuclear Instruments and Methods in Physics Research A {\bf 499}, 001 007 (2003), and other papers included in the volume; T.~Abe \textit{et al.}, Prog. Theor. Exp. Phys. {\bf 2013}, 03A001 (2013) and following articles up to 03A011.

\bibitem{detector}
A.~Abashian {\it et al.}, Nuclear Instruments and Methods in Physics Research A {\bf 479}, 117 232 (2002); also see Section 2 in J.~Brodzicka {\it et al.}, Prog. Theor. Exp. Phys. {\bf 2012}, 04D001 (2012).

\bibitem{evtgen}
D.J.~Lange, Nucl. Instrum. Methods Phys. Res. Sect. A {\bf 462}, 152 (2001).

\bibitem{photos}
E.~Barberio and Z.~Wąs, Comput. Phys. Commun. {\bf 79}, 291 (1994).

\bibitem{tauola}
S.~Jadach, Z.~Was, R.~Decker, and J. H.~Kuhn, Comp. Phys. Commun. {\bf 76}, 361 (1993).

\bibitem{pythia}
T.~Sjostrand, S.~Mrenna, and P.~Skands, JHEP {\bf 0605}, 026 (2006).

\bibitem{geant}
R.~Brun {\it et al.}, GEANT3, CERN-DD-EE-84-1 (1987).

\bibitem{pid}
E.~Nakano, Nucl. Instrum. Methods Phys. Res. Sect. A {\bf 494}, 402 (2002).

\bibitem{muid}
A.~Abashian {\it et al.}, Nucl. Instrum. Meth. Phys. Res., Sect. A {\bf 491}, 69 (2002).

\bibitem{eid}
K.~Hanagaki {\it et al.}, Nucl. Instrum. Meth. Phys. Res., Sect. A {\bf 485}, 490 (2002).

\bibitem{pdg}
P.A.~Zyla {\it et al.} (Particle Data Group), Prog. Theor. Exp. Phys. {\bf 2020}, 083C01 (2020) and 2021 update.

\bibitem{ny2s}
X.L.~Wang {\it et al.} (Belle Collaboration), Phys. Rev. D {\bf 84}, 071107(R) (2011).

\bibitem{t2lg}
K.~Uno {\it et al.} (Belle Collaboration), J. High Energ. Phys. {\bf 2021}, 19 (2021). 

\bibitem{ul}
S.~Sandilya {\it et al.} (Belle Collaboration), Phys. Rev. D {\bf 98}, 071101 (2018).

\end{thebibliography}
\end{document}